\documentclass[12pt]{article}
 \usepackage{epsfig}
 \def\be{\begin{equation}}
 \def\ee{\end{equation}}
 \def\bea{\begin{eqnarray}}
 \def\eea{\end{eqnarray}}
 \usepackage{graphicx}

 \catcode`\@=11
 \def\lsim{\mathrel{\mathpalette\@versim<}}
 \def\gsim{\mathrel{\mathpalette\@versim>}}
 \def\@versim#1#2{\vcenter{\offinterlineskip
 \ialign{$\m@th#1\hfil##\hfil$\crcr#2\crcr\sim\crcr } }}
 \catcode`\@=12

 \catcode`@=12
\begin{document}
 \thispagestyle{empty}
 \begin{flushright}
 UCRHEP-T575\\
 February 2017\
 \end{flushright}
 \vspace{0.6in}
 \begin{center}
 {\Large \bf The Many Guises of a Neutral Fermion Singlet\\}
 \vspace{1.2in}
 {\bf Ernest Ma\\}
 \vspace{0.2in}
 {\sl Physics and Astronomy Department,\\ 
 University of California, Riverside, California 92521, USA\\}
 \end{center}
 \vspace{1.2in}

\begin{abstract}\
The addition of a neutral fermion singlet to the standard model of particle 
interactions leads to many diverse possibilities.  It is not necessarily 
a right-handed neutrino.  I discuss many of the simplest and most 
interesting scenarios of possible new physics with this approach.  
In particular I propose the possible spontaneous breaking 
of baryon number, resulting in the massless 'sakharon'.
\end{abstract}

 \newpage
 \baselineskip 24pt

\noindent \underline{\it Introduction}~:~\\
In the standard model (SM) of particle interactions, there are three families 
of quarks and leptons.  Under its $SU(3)_C \times SU(2)_L \times U(1)_Y$ 
gauge symmetry, they transform as follows:
\begin{eqnarray}  
&& \pmatrix{u \cr d}_L \sim (3,2,1/6), ~~~ u_R \sim (3,1,2/3), ~~~ 
d_R \sim (3,1,-1/3), \\ 
&& \pmatrix{\nu \cr l}_L \sim (1,2,-1/2), ~~~ l_R \sim (1,1,-1), 
\end{eqnarray}
where electric charge is given by
\begin{equation}
Q = I_{3L} + Y.
\end{equation}
There is also one scalar Higgs doublet
\begin{equation}
\Phi = \pmatrix{\phi^+ \cr \phi^0} \sim (1,2,1/2),
\end{equation}
which breaks the electroweak $SU(2)_L \times U(1)_Y$ symmetry spontaneously 
to electromagnetic $U(1)_Q$ through the vacuum expectation value $\langle 
\phi^0 \rangle = v$.   As a result, three vector gauge bosons 
$W^\pm,Z$ become massive, whereas the eight $SU(3)_C$ gluons and the one 
$U(1)_Q$ photon remain massless.  There is also just one physical real 
scalar, i.e. the Higgs boson, presumably the 125 GeV discovered in 2012 
at the Large Hadron Collider (LHC)~\cite{atlas12,cms12}.

As it stands, the standard model has the following automatic conserved 
global symmetries: baryon number $B=1/3$ for each quark,  lepton number 
$L_e = 1$ for 
the electron and its neutrino $\nu_e$,  $L_\mu = 1$ for $\mu$ and $\nu_\mu$, 
and $L_\tau = 1$ for $\tau$ and $\nu_\tau$.  As such, all neutrinos are 
massless.

Because of the observation of neutrino oscillations, we know that at least 
two neutrinos are massive, and that $\nu_e, \nu_\mu, \nu_\tau$  are not 
mass eigenstates.  The simplest theoretical implementation of this fact 
is to add one singlet neutral fermion $N_R$ to each family:
\begin{equation}
N_R \sim (1,1,0).
\end{equation}
Because of Eq.~(3), $N_R$ has no gauge interaction and it couples to the 
SM only through the Yukawa terms
\begin{equation}
f^\nu_{ij} \bar{N}_{iR} (\nu_{jL} \phi^0 - l_{jL} \phi^+) + H.c.
\end{equation}
Hence $N_R$ is commonly called the right-handed neutrino and assigned lepton 
number $L=1$ with its Dirac mass matrix given by $f^\nu_{ij}v$.  The separate 
conservation of $L_e, L_\mu, L_\tau$ is no longer valid, replaced now 
by $L = L_e + L_\mu + L_\tau$.

If the Majorana mass terms
\begin{equation}
{1 \over 2} M_{ij} N_{iR} N_{jR} + H.c.
\end{equation}
are added, then $L$ is broken to $(-1)^L$, and for large $M_{ij}$, the 
light neutrino mass matrix is given by the famous seesaw 
formula~\cite{m77,y79,gmrs79,ms80}
\begin{equation}
{\cal M}^\nu_{ij} = - (f^\nu_{ik} v) M^{-1}_{kl} (f^\nu_{lj} v).
\end{equation}

\noindent \underline{\it Lepton number extensions}~:~\\
Suppose $N_R$ is still assumed to have $L=1$ as implied by Eq.~(6), and 
that Eq.~(7) is forbidden, then another interesting possibility exists 
if a scalar singlet $\sigma$ with $L=-2$ is added.  Now the terms
\begin{equation}
{1 \over 2} f^N_{ij} \sigma N_{iR} N_{jR} + H.c.
\end{equation}
would generate a mass matrix $f^N_{ij} \langle \sigma \rangle$.  However, 
the spontaneous breaking of $L$ implies a massless Goldstone boson, i.e. 
the singlet majoron~\cite{cmp81}. 

The above model can be interpreted another way if we add a heavy 
singlet quark~\cite{k79,svz80} which also transforms under $L$.  However, 
whereas $Q_L$ has $L=1$, its chiral partner $Q_R$ has $L=-1$.  In that case, 
the Yukawa term
\begin{equation}
f^Q \sigma \bar{Q}_R Q_L + H.c.
\end{equation}
exists and $L$ becomes an anomalous global symmetry which is identifiable 
with the Peccei-Quinn symmetry~\cite{pq77} which solves the strong CP 
problem, and a very light 
axion~\cite{wb78,wc78} appears instead of the massless majoron.  
This idea~\cite{s87} also connects the axion scale with the 
neutrino mass seesaw scale, and may be extended~\cite{m01-1} to include 
supersymmetry.

Since $N_R$ is a new addition to the SM, we are free to assign it whatever 
symmetry we desire.  Suppose it has $L=0$.  This means that Eq.~(7) is 
allowed, but Eq.~(6) is forbidden, and there is no connection between 
$N_R$ and the SM.  However, suppose we now add a second scalar doublet
\begin{equation}
\eta = \pmatrix{\eta^+ \cr \eta^0} \sim (2,1 ,1/2)
\end{equation}
with $L=-1$, then the terms
\begin{equation}
f^\eta_{ij} \bar{N}_{iR} ( \nu_{jL} \eta^0 - l_{jL} \eta^+) + H.c.
\end{equation}
are allowed, and if $L$ is spontaneously broken by $\langle \eta^0 \rangle 
= u$, neutrinos would become massive, but a massless doublet majoron would 
also appear.  It would contribute to the invisible decay width of the $Z$ 
boson, which is known to be consistent with exactly three neutrinos. 
This scenario is thus ruled out.

Suppose now that $L$ is also broken explicitly but softly by the 
bilinear term
\begin{equation}
\mu^2 \eta^\dagger \Phi + H.c.,
\end{equation}
then it can be shown~\cite{m01} that $u << v$ naturally, and the smallness 
of the neutrino Dirac mass $f^\eta u$ is understood.  Together with the 
seesaw mechanism of Eq.~(8), this implies that the mass of $N_R$ may be 
reduced to below 1 TeV, lending hope that the seesaw mechanism may be 
verifiable experimentally.

Another possible lepton number assignment for $N_R$ is $L=-1$.  Now both 
Eqs.~(6) and (7) are forbidden.  Suppose we add $\eta$ but with $L=-2$ 
instead, then Eq.~(12) is allowed.  Using again the dimension-two term 
of Eq.~(13) to break $L$ softly, we obtain Dirac masses for the light 
neutrinos~\cite{dl09,mp17} without the dimension-three Majorana mass terms of 
Eq.~(7).  The resulting Lagrangian actually conserves the usual $L$.  
What has been gained is the understanding of how Dirac neutrino masses 
may be small at the expense of a second scalar doublet with a suppressed 
vacuum expectation value.

\noindent \underline{\it Dark matter extensions}~:~
Another possible identity for $N_R$ is dark matter (DM).  Suppose it is odd 
under an exactly conserved $Z_2$ discrete symmetry under which all SM 
particles are even.  This scenario is actually identical to the case 
$L=0$ discussed in the previous section, i.e. Eq.~(6) is forbidden but 
Eq.~(7) is allowed.  They are related by defining dark matter parity 
\begin{equation}
D = (-1)^{L+2j},
\end{equation} 
as pointed out in Ref.~\cite{m14}.  However, $N_R$ 
decouples entirely from the SM and may not be relevant as a DM candidate. 

To connect $N_R$ to the SM, the scalar doublet $\eta$ of Eq.~(11) may again 
be added, and the Yukawa couplings of Eq.~(12) be allowed by assigning 
$\eta$ to be odd under dark $Z_2$.  Now the concept of lepton number 
for Eq.~(12) becomes ambiguous.  It could be assigned to $N_R$ or $\eta$.  
However, if we abandon $L$ and just consider lepton parity, i.e. $(-1)^L$, 
with $L=0$ for $N_R$ and $L=-1$ for $\eta$, then this term conserves 
both lepton parity and dark parity.  In fact the latter is 
derivable~\cite{m15} from the former as shown in Eq.~(14).

With both $N_R$ and $\eta$, it is now possible to obtain a radiative seesaw 
neutrino mass~\cite{m06}, as shown in Fig.~1.
\begin{figure}[htb]
\vspace*{-3cm}
\hspace*{-3cm}
\includegraphics[scale=1.0]{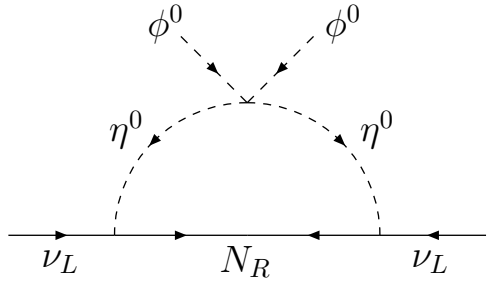}
\vspace*{-21.5cm}
\caption{One-loop $Z_2$ scotogenic seesaw neutrino mass.}
\end{figure}
Since this mechanism uses dark matter to generate a nonzero neutrino mass, 
it has been called 'scotogenic' from the Greek 'scotos' meaning darkness.
The concept of lepton parity may be promoted to matter parity
\begin{equation}
M = (-1)^{3B+L},
\end{equation}
so that dark matter parity becomes
\begin{equation}
D = (-1)^{3B+L+2j},
\end{equation}
which is identical to that of $R$ parity in supersymmetry, but without 
having to extend the SM to include supersymmetry itself.  In that case, 
with the addition of scalar fermion doublets and singlets of odd 
$D$ parity, all quark and lepton masses may be generated~\cite{m14} 
from the heavy $3 \times 3$ $N_R$ Majorana mass matrix.  This idea 
connects SM masses to those of the dark sector, and offers an explanation 
for the light fermion masses as being scotogenic.

\noindent \underline{\it Gauge U(1) extensions}~:~
Whereas $N_R$ has no SM gauge interactions, it may transform nontrivially 
under an extra gauge U(1) symmetry.  The most commonly studied scenario 
is gauge $B-L$ from the observation that~\cite{mm86}
\begin{equation}
Q = I_{3L} + I_{3R} + {1 \over 2} (B-L)
\end{equation}
for the known quarks and leptons so that the SM may be embedded into 
$SU(3)_C \times SU(2)_L \times SU(2)_R \times U(1)_{B-L}$.  On the other 
hand, if we consider $SU(3)_C \times SU(2)_L \times U(1)_Y \times U(1)_F$ 
with one $N_R$ added to each family quarks and leptons as shown in 
Table 1, then many possible different models~\cite{kmpz17} may be obtained.
\begin{table}[htb]
\caption{Fermion assignments under $U(1)_F$.}
\begin{center}
\begin{tabular}{|c|c|c|c|c|}
\hline
Particle & $SU(3)_C$ & $SU(2)_L$ & $U(1)_Y$ & $U(1)_F$ \\
\hline
$Q_{iL} = (u,d)_{iL}$ & 3 & 2 & 1/6 & $n_i$ \\
$u_{iR}$ & $3$ & 1 & $2/3$ & $n_i$ \\
$d_{iR}$ & $3$ & 1 & $-1/3$ & $n_i$ \\
\hline
$L_{iL} = (\nu,l)_{iL}$ & 1 & 2 & $-1/2$ & $n'_i$ \\
$l_{iR}$ & 1 & 1 & $-1$ & $n'_i$ \\
$N_{iR}$ & 1 & 1 & 0 & $n'_i$ \\
\hline
\end{tabular}
\end{center}
\end{table}
To constrain $n_{1,2,3}$ and $n'_{1,2,3}$, the requirement of gauge anomaly 
cancellation is imposed.  The contributions of color triplets to the 
$[SU(3)]^2 U(1)_F$ anomaly sum up to
\begin{equation}
[SU(3)]^2 U(1)_F~: ~~~ {1 \over 2} \sum_{i=1}^3 (2n_i - n_i - n_i);
\end{equation}
and the contributions of $Q_{iL},u_{iR},d_{iR},L_{iL},l_{iR}$ to the 
$U(1)_Y [U(1)_F]^2$ anomaly sum up to
\begin{equation}
U(1)_Y [U(1)_F]^2~: ~~~ \sum^3_{i=1} \left[ 6 \left( {1 \over 6} \right) - 
3 \left( {2 \over 3} \right) - 3 \left( -{1 \over 3} \right) \right] n_i^2 
+ \left[ 2 \left( -{1 \over 2} \right) - (-1) \right] {n'_i}^2.
\end{equation}
Both are automatically zero, as well as the $[U(1)_F]^3$ anomaly because 
all fermions couple to $U(1)_F$ vectorially.  The contributions of the 
$SU(2)_L$ doublets to the $[SU(2)]^2 U(1)_F$ anomaly sum up to
\begin{equation}
[SU(2)]^2 U(1)_F~: ~~~ {1 \over 2} \sum^3_{i=1} (3n_i + n'_i);
\end{equation}
and the contributions to the $[U(1)_Y]^2 U(1)_F$ anomaly sum up to
\begin{eqnarray}
[U(1)_Y]^2 U(1)_F &:&  \sum^3_{i=1} \left[ 6 \left( {1 \over 6} \right)^2 - 
3 \left( {2 \over 3} \right)^2 - 3 \left( -{1 \over 3} \right)^2 \right] n_i 
+ \left[ 2 \left( -{1 \over 2} \right)^2 - (-1)^2 \right] n'_i \nonumber \\ 
&=& \sum^3_{i=1} \left( -{3 \over 2} n_i - {1 \over 2} n'_i \right).
\end{eqnarray}
Both are zero if
\begin{equation}
\sum^3_{i=1} (3n_i + n'_i) = 0.
\end{equation}
There are many specific examples of models which satisfy this condition as 
dicussed in Ref.~\cite{kmpz17}.

The neutral vector gauge boson $Z_F$ associated with $U(1)_F$ couples in 
general to the $u$ and $d$ quarks, so it may be produced at the LHC 
if kinematically allowed.  Its branching fractions to $e^-e^+$ and 
$\mu^- \mu^+$ are given by
\begin{equation}
B(Z_F \to e^-e^+,\mu^-\mu^+) = {2 {n'_{1,2}}^2 \over 12 \sum n_i^2 + 3 \sum 
{n'_i}^2}.
\end{equation}
The $c_{u,d}$ coeffficients used in the experimental 
search~\cite{atlas14,cms15} of $Z_F$ are then
\begin{equation}
c_u = c_d = 2 n_1^2 g_F^2 (2 {n'_1}^2 + 2 {n'_2}^2)/(12 \sum n_i^2 + 3 \sum 
{n'_i}^2).
\end{equation}
With current LHC data, a typical bound~\cite{kmpz17} on $Z_F$ is about 4 TeV.

\noindent \underline{\it Baryon number extensions}~:~
An interesting but seldom explored possibility is to make $N_R$ a baryon. 
Suppose a scalar quark
\begin{equation}
\zeta \sim (3,1,-1/3)
\end{equation}
is added with $B = -2/3$ so that the Yukawa terms
\begin{equation}
{f_L^\zeta}_{ij} \zeta (d_{iL} u_{jL} - u_{iL} d_{jL}) +{f_R^\zeta}_{ij} \zeta 
d_{iR} u_{jR} + f^N_{ij} N_{iR} d_{jR} \zeta^* + H.c.
\end{equation}
are allowed, then $N_R$ has $B=-1$.  This assignment was first 
proposed~\cite{m88} in the context of superstring-inspired $E_6$ models. 
Note that both Eqs.~(6) and (7) are forbidden by $B$, but if the latter 
is allowed, then $B$ is broken softly to $B$ parity.  In that case, 
whereas proton decay is still forbidden, neutron-antineutron ($n - \bar{n}$) 
oscillation is possible.

The scotogenic mechanism may also be extended to accommodate~\cite{gms16} 
$n - \bar{n}$ oscillation.  The idea is very simple.  Replace $\Phi$ by 
the singlet scalar quark $\delta \sim (3,1,-1/3;+)$ and $\eta$ by 
$\xi \sim (3,1,-1/3;-)$, which is distinguished from $\delta$ by having 
odd $B$ parity. Together with $N_R$ having even $B$ parity, Fig.~1 becomes  
Fig.~2.
\begin{figure}[htb]
\vspace*{-3cm}
\hspace*{-3cm}
\includegraphics[scale=1.0]{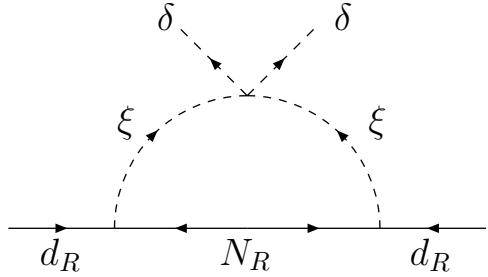}
\vspace*{-21.5cm}
\caption{One-loop $Z_2$ scotogenic $n - \bar{n}$ oscillation.}
\end{figure}
Since $\delta$ acts as a diquark because it couples to $u_{L,R} d_{L,R}$, 
this diagram generates $n - \bar{n}$ oscillation.  The idea of combining 
Figs.~1 and 2 means that neutrino mass, $n - \bar{n}$ oscillation are 
possible only through their connection to dark matter.  Proton decay 
is forbidden at this stage by the separate conservation of $L$ parity and 
$B$ parity, but if only the product is conserved, then it also 
becomes possible~\cite{gms16}.

Another use of $N_R$ having $B=1$ is in the context of supersymmetry.  
Since the fermionic component of the $N_R$ superfield has even $R$ parity, 
whereas the bosonic component has odd $R$ parity, the latter may be 
dark matter.  The addition of a pair of color triplet superfields with 
weak hypercharge $= \pm 2/3$ or $\mp 1/3$ may also facilitate baryogenesis 
from the decay of the $N_R$ fermion at the TeV scale~\cite{bmn07}. 

Instead of having an allowed mass term for $N_R$ as in Eq.~(7) so that 
$B$ is broken to $B$ parity in the above, we can generalize the idea of 
the spontaneous breaking of lepton number to that of baryon number. 
The resulting massless Goldstone boson may be called the 'sakharon', 
after Andrei Sakharov.  To implement this idea in a renormalizable 
extension of the Standard Model, the simplest solution is 
to use Eq.~(26) and add Eq.~(9) instead of Eq.~(7). 
Hence $\zeta$ decays into $\bar{u} \bar{d}$ and $\zeta^*$ decays into $u d$.  
As $\sigma$ acquires a large vacuum expectation value, 
$B$ is broken to $(-)^{3B}$ and $N$ may decay into $udd$ or $\bar{u} 
\bar{d} \bar{d}$.  If there are two or more $N$ fields, this is 
a mechanism for generating a baryon asymmetry~\cite{s67} in the early 
Universe, which gets converted into a $B-L$ asymmetry through the 
electroweak sphalerons~\cite{dnn08}.

In this scenario where $B$ is spontaneously broken at a very high scale, 
the sakharon $S$ couples directly only to $N$, just as in the case of 
the singlet majoron.  However, whereas $N$ would mix with $\nu$ 
in the presence of electroweak symmetry breaking, it stands alone in this 
scenario. 
This means that there is no tree-level sakharon interaction with ordinary 
matter, and its presence is even more elusive than that of the singlet 
majoron~\cite{cmp80}.

Consider now the extreme opposite scenario of a very low energy scale for the 
spontaneous breakdown of baryon number.  This is somewhat akin to the case 
of the triplet majoron model~\cite{gr81} where lepton number is spontaneously 
broken at a very low scale, i.e. that of neutrino mass.  That is however 
experimentally ruled out because the triplet majoron interacts with the 
$Z$ boson, which decays into it and its necessarily light scalar partner 
so that the $Z$ invisible width is increased by twice that of a single 
neutrino species.  Here the sakharon will be a singlet as detailed below.

\begin{table}[htb]
\caption{Particle content of model with low-scale sakharon.}
\begin{center}
\begin{tabular}{|c|c|c|c|}
\hline 
Particle & $SU(3)_C \times SU(2)_L \times U(1)_Y$ & $B$ & $L$ \\ 
\hline
$Q = (u,d)_L$ & (3,2,1/6) & 1/3 & 0 \\
$u_R$ & $(3,1,2/3)$ & 1/3 & 0 \\ 
$d_R$ & $(3,1,-1/3)$ & 1/3 & 0 \\ 
\hline
$h_L$ & $(3,1,-1/3)$ & $-2/3$ & $-1$ \\ 
$h_R$ & $(3,1,-1/3)$ & $-2/3$ & $-1$ \\ 
\hline
$L = (\nu,l)_L$ & $(1,2,-1/2)$ & 0 & 1 \\ 
$l_R$ & $(1,1,-1)$ & 0 & 1 \\
$N_R$ & (1,1,0) & 0 & 1\\ 
\hline
\hline
$(\phi^+,\phi^0)$ & $(1,2,1/2)$ & 0 & 0 \\ 
\hline
$\zeta$ & $(3,1,-1/3)$ & $-2/3$ & 0 \\
\hline
$\sigma$ & $(1,1,0)$ & 1 & 1 \\
\hline
\end{tabular}
\end{center}
\end{table}
To implement this extreme scenario, the Standard Model of quarks and 
leptons is extended to include three heavy Majorana singlet neutral fermions 
$N_R$ (for obtaining small neutrino masses through the canonical seesaw 
mechanism) together with singlet quarks $h_{L,R}$, as well as a color-triplet 
scalar $\zeta$ and a complex singlet scalar $\sigma$.  Their baryon and 
lepton numbers are listed in Table 2.  The interaction Lagrangian is 
then given by
\begin{equation}
{\cal L}_{int} = {f_L^\zeta}_{ij} \zeta (d_{iL} u_{jL} - u_{iL} d_{jL}) 
+ {f_R^\zeta}_{ij} \zeta d_{iR} u_{jR} + f^N_k \zeta^* N_{kR} h_R + 
f^\sigma \bar{d}_{kR} h_L \sigma + H.c.
\end{equation}
Allowing $N_R$ to have a large Majorana mass breaks $L$ to $(-)^L$, 
under which ${\cal L}_{int}$ of Eq.~(27) is still invariant.  Consider now the 
possibility of $\langle \sigma \rangle \neq 0$, thereby breaking both $B$ and 
$(-)^L$.  Although $(-)^{3B+L}$ remains unbroken in this case, it does not 
impose any extra condition because all fermions are odd and all bosons are 
even under it.  There are many consequences of this scenario.  Proton decay 
is now possible, but is suppressed by two factors: the smallness of 
$\langle \sigma \rangle$ and the smallness of $m_\nu$.  Details will be 
reported elsewhere~\cite{gms17}.

\noindent \underline{\it Conclusion}~:~
In this Brief Review, some of the many guises of a neutral fermion singlet 
are exposed.  In its simplest form, it is used as a right-handed neutrino, 
but many other options are available.  I have discussed lepton and baryon 
number extensions, axion and dark matter applications, as well as gauge 
$U(1)$ family symmetries.  The lesson is that for any new particle added 
to the SM, its lepton or baryon number assignment has to be understood in 
context, and not an automatic entry.

\noindent \underline{\it Acknowledgement}~:~
This work was supported in part by the U.~S.~Department of Energy Grant 
No. DE-SC0008541.

\bibliographystyle{unsrt}

\end{document}